# Efficient quasi-monoenergetic ion beams up to 18 MeV/nucleon via self-generated plasma fields in relativistic laser plasmas

Sasi Palaniyappan[*,1], Chengkun Huang[1], Donald C. Gautier[1], Christopher E. Hamilton[1], Miguel A. Santiago[1], Christian Kreuzer[2], Rahul C. Shah[1], and Juan C. Fernández[1].

[1]Los Alamos National Laboratory, Los Alamos, New Mexico 87545, USA.
[2]Ludwig-Maximilian-University, Munich, Germany.

*e-mail: sasi@lanl.gov

**Table-top laser-plasma ion accelerators seldom achieve narrow energy spreads, and never without serious compromises in efficiency, particle yield, etc. Using massive computer simulations, we identify a self-organizing scheme that exploits persisting self-generated plasma electric (~TV/m) and magnetic (~$10^4$ Tesla) fields to reduce the ion energy spread after the laser exits the plasma – separating the ion acceleration from the energy spread reduction. Consistent with the scheme, we experimentally demonstrate aluminum and carbon ion beams with narrow spectral peaks at energies up to 310 MeV (11.5 MeV/nucleon) and 220 MeV (18.3 MeV/nucleon), respectively, with high conversion efficiency (~5%, i.e., 4J out of 80J laser). This is achieved with 0.12 PW high-contrast Gaussian laser pulses irradiating planar foils with optimal thicknesses of up to 250 nm that scale with laser intensity. When increasing the focused laser intensity fourfold (by reducing the focusing optic f/number twofold), the spectral-peak energy increases twofold. These results pave the way for next generation compact accelerators suitable for applications. For example, 400 MeV (33.3 MeV/nucleon) carbon-ion beam with narrow energy spread required for ion fast ignition could be generated using PW-class lasers.**

## Introduction

Laser-driven ion beams with narrow energy spread and high conversion efficiency would be transformational because they can deliver unprecedented power densities. For example, warm dense matter with conditions relevant to stars and planetary cores can be created in the laboratory by isochoric heating of bulk matter with such ion beams [1, 2]. Ion fast ignition is an extreme example of isochoric heating, where laser-driven ion beams can ignite compressed fuel to generate fusion energy [3-7].

Despite a decade-plus effort, achieving laser-driven ion beams with simultaneous narrow energy spread and high efficiency is still challenging [8-20]. Widely explored schemes for such ion beam generation include target-normal-sheath-acceleration (TNSA) [9-11, 21-29], radiation-pressure-acceleration (RPA) [12-14, 20, 30-32], breakout-afterburner (BOA) [33-35], magnetic vortex acceleration[36-39], and collisionless-shock-acceleration [16, 17]. In particular, Ref. [14] reported ~80 MeV $C^{6+}$ ion beam with ~70% energy spread and ~1% conversion efficiency using a 0.25 PW laser.

Here we report a self-organizing scheme where persistent quasi-static plasma electric and magnetic fields reduce the large ion-energy spread after the laser-plasma interaction ends. This scheme largely separates the energy-spread reduction from the acceleration phase. Thus, we experimentally demonstrate laser-driven ion beams with narrow spectral peaks at energies up to 310 MeV for $Al^{11+}$ and 220 MeV for $C^{6+}$ and high conversion efficiency (4-5%). Also, we demonstrate that increasing the focused laser intensity fourfold (by reducing the focusing optic f/number twofold), the spectral-peak energy increases twofold. These results are obtained with 0.12 PW (80J, 650-fs, linear polarized) high-contrast Gaussian laser pulses from the Trident laser [40] at the Los Alamos National Laboratory, irradiating planar foils of an optimized thickness of up to 250 nm (see below).

## Ion acceleration & spectral shaping scheme

The scheme reported here, depicted in Fig. 1, proceeds through these basic steps. (i) Initially, the laser interacts with an opaque overdense plasma target, which reflects much of it (Fig. 1a). The laser-heated front side of the plasma expands towards the laser, while the laser-driven hot electrons heat the plasma interior and the rear surface expands away from the laser (TNSA process). Compensating transverse return electron currents flow towards the laser focal region generating azimuthal magnetic fields (Fig. 1a).

(ii) As the plasma heating/expansion continues, the laser drives the plasma electrons to near-light speed, which makes them heavier and slower to respond to the laser. This enables laser penetration into the opaque plasma slab, i.e. "relativistic transparency" [41]. As in previous work, timing of relativistic transparency to get a volumetric laser-plasma interaction and to avoid premature target expansion requires an optimal target thickness versus intensity [33-35]. Subsequently, the laser drives a large forward electron current and associated azimuthal magnetic field ($\sim 10^4$ T) strong enough to magnetize the electrons, but not the ions (Fig. 1b). (Self-generated magnetic fields have been reported in previous studies of laser-driven plasmas [36-38, 42-47]). The laser-driven forward plasma flow is largely confined within a central channel. Although the B-field acts like a funnel that restricts the forward electron flow, the channel still has higher charge density of electrons than ions as the laser vigorously drives the electrons into the channel against the magnetic resistance. At this time, the ions within the plasma channel exhibit an exponential energy spectrum and they are "chirped" in space with the fastest (slowest) ones ahead (behind).

(iii) After the laser exits the plasma, the electrons' forward motion within the channel slows down as they interact with the B-field, without the laser drive (Fig. 1c). Additionally, the plasma channel could be kinked (see simulation Fig. 5c&h), adding more resistance to the electron's forward motion and spatially bunching them. While the channel electron population is leaking forward, it is partially replenished by the remaining quasi-static longitudinal electric field (fig. 1c) at the front side of the plasma (from the earlier plasma expansion towards the laser), albeit not as strongly as with the laser on.

Meanwhile, the ions continue to move forward unaffected by the magnetic field. This leads to a net positive charge around the localized electron bunch (Fig. 1d). Such a charge distribution will rearrange itself by accelerating the (lagging) slower ions and decelerating the (leading) faster ions, creating an ion spectral peak (Fig. 1d).

The late-time dynamics mediated by self-generated plasma fields explored here are complementary (not contradictory) to earlier work with BOA and ion acceleration in the relativistic transparency regime [33-35]. BOA refers to a brief period of enhanced ion acceleration upon the onset of relativistic transparency. Previous BOA simulations have not shown persistent ion spectral peaks. Here, we discuss how optimization of the laser-plasma interaction in the relativistic transparency regime mediated by self-generated plasma fields could lead to ion spectral bunching after the laser exits the plasma.

## Results

| Laser focus | Intensity (W/cm$^2$) | Target[†] | Plasma expansion speed towards laser (μm/ps)* | Ion spectral peak[¶] | FWHM Energy spread[¶] (%) | Conversion efficiency (%) | Proton spectral peak[¶] (MeV) |
|---|---|---|---|---|---|---|---|
| f/3 | 2 x 10$^{20}$ | 110nm Al | 1.7 | 165 MeV Al$^{11+}$ (6.1 MeV/u) | 7 | 5 | Below TP cut-off |
| f/3 | 2 x 10$^{20}$ | 100nm C | 0.6[‡] | No spectral peak[§] | - | - | No spectral peak[§] |
| f/3 | 2 x 10$^{20}$ | 110nm Al /10nm C | - | 80 MeV C$^{6+}$ (6.7 MeV/u) | 15 | - | Below-TP cut-off |
| f/1.5 | 8 x 10$^{20}$ | 250nm Al | 1.1 | 310 MeV Al$^{11+}$ (11.5 MeV/u) | 41 | 4 | 12.6 |
| f/1.5 | 8 x 10$^{20}$ | 250nm Al /10nm C | - | 120 MeV C$^{6+}$ (10 MeV/u) | 54 | - | 17 |
| f/1.5 | 8 x 10$^{20}$ | 250nm C | -0.46 | No spectral peak | - | - | No spectral peak |
| f/1.5 | 6 x 10$^{20}$ | 250nm C | 1.2 | 220 MeV C$^{6+}$ (18.3 MeV/u) | 23 | 4 | 23.3 |

FWHM – full width at half maximum; TP – Thomson parabola ion energy diagnostic
* Obtained from the frequency shift of the laser light reflected from the plasma before the onset of relativistic transparency using single-shot FROG (Frequency-Resolved-Optical-Gating)
[†] 'a/b' notation means the target 'b' is coated on the rear side (i.e., away from the incoming laser) of target 'a'
[¶] Measured at 0° on-axis with the laser beam; First number denotes ion spectral peak. Number in parenthesis denotes the same in units of energy per nucleon. [‡]Obtained from Ref. [41]; [§] Obtained from Ref. [34]; - not discussed in the article

**Table 1: laser/target parameters and properties of generated ion beams and plasmas**

Table 1 summarizes the laser/target parameters along with the resulting plasma and ion beam properties discussed in this article.

**Generation of 165 MeV Al$^{11+}$ & 80 MeV C$^{6+}$ ion beams**

Fig. 2 shows schematic layout of the experiment. An f/3 off-axis-parabola focuses the 0.12 PW Trident laser (2 x 10$^{20}$ W/cm$^2$ peak intensity) onto a 110 nm thick aluminum foil (density - 2.7 g/cm$^3$). Multiple optical and particle diagnostics characterized the experiment (Table 1 – second row). The raw Thomson parabola (TP) data (Fig. 2a) shows dominant Al$^{11+}$ and proton impurity traces along with a faint trace of Al$^{12+}$. Atomic ionization calculation of aluminum for Trident laser parameters (f/3 focus) shows the aluminum is ionized to Al$^{11+}$ and there is barely any Al$^{12+}$ due to the huge inner-shell gap in the ionization potential of Al$^{11+}$ and Al$^{12+}$ (see Supplementary Fig. 1) [48]. The Al$^{11+}$ ion energy spectrum measured on-axis (Fig. 2b) peaks at 166 MeV (i.e., 6.2 MeV/nucleon) with a 7% energy spread. The integrated spectrum yields a total of 1.8x10$^9$ ions per millisteradian (msr) with average energy of 123 MeV. The corresponding spectrum measured at 8.5° off-axis (Fig. 2c), for another shot with the same target and laser conditions, shows a similar ion peak at 165 MeV with 30% energy spread (total of 5.5x10$^8$ ions/msr with average energy of 131 MeV). The proton spectrum has no spectral peak from 9 MeV up to 50 MeV (see Supplementary Fig. 2). TP settings precluded measuring protons below 9 MeV.

Angularly resolved ion energy spectra (Fig. 2e) show aluminum ions up to 34° Full-Width-Half-Maximum. From the measured ion beam divergence, assuming radial symmetry, we estimate a total of 2 x 10$^{11}$ Al$^{11+}$ ions in the full beam (average energy of 127 MeV). This implies 4J in the aluminum ion beam out of the 80J incident laser energy (5% conversion efficiency) and 0.35 µC of charge. A similar calculation for protons yields 0.6% conversion efficiency (see Supplementary Fig. 2).

The sharp drop in reflected laser-light intensity (Fig. 2h) and the temporal phase reversal of the transmitted laser light (Fig. 2i) are indicative of the ensuing relativistic transparency phase consistent with earlier results [41]. The transmitted laser beam (Fig. 2d) contains 20% of the incident laser energy i.e., 16J. Relativistic transparency enables strong laser coupling into the dense plasma driving stronger forward electron current and magnetic fields.

Similar laser interaction with synthetic diamond foils did not show spectrally peaked carbon ion beam (Table 1 – third row)[34]. This is because the diamond foil pre-expanded from the Trident laser pedestal significantly more than the aluminum foil, resulting in lower initial peak plasma density as indicated by the reflected laser spectrum and FROG measurements. When the main laser pulse initially heats the aluminum foil, it expands towards the laser, and the reflected light is Doppler blue-shifted. The time-integrated reflected light spectrum (Fig. 2g) from the aluminum foil shows a 9 nm blue-shifted peak at 1044 nm ($d\lambda/\lambda = 2v/c \cong 0.009$), i.e., a plasma expansion of 1.4 µm/ps towards the incoming laser. The time-resolved temporal phase of the reflected light from aluminum (Fig. 2h) shows an early frequency blue-shift $d\omega/\omega = 0.011$, i.e., a plasma expansion of 1.7 µm/ps towards the laser. The reflected laser from the diamond does not show any noticeable blue-shifted peak (see Fig. 4b inset in ref. [41]). Also, the reflected laser from diamond (see fig. 4b in ref. [41]) shows an early frequency blue-shift of $d\omega/\omega = 0.004$ (plasma expansion of 0.6 µm/ps towards the incoming laser) – nearly three times slower expansion compared to aluminum. Perhaps counterintuitive, the slower plasma expansion towards the laser means the initial peak plasma density is lower and the laser better able to push it forward.

This conclusion is supported by HELIOS rad-hydro[49] simulations of aluminum and diamond foils pre-expansion. According to the simulations (Fig. 1f), the peak carbon density (0.8 g/cm$^3$) is three times lower than the peak aluminum density (2.4 g/cm$^3$) when expanded by the same Trident laser pedestal (see supplementary Fig. 3 for Trident contrast). Lower initial plasma density leads to lower residual self-generated plasma fields, too weak to reduce the ion energy spread after the laser-plasma interaction ends – a hypothesis consistent with particle-in-cell simulations discussed later in this article.

The contribution of hole-boring radiation-pressure (HB-RPA) ion acceleration to the observed ion energy peak is minimal. The maximum spectral red-shift of the reflected light in Fig. 2g is ~50 nm as the plasma moves away from the laser, which yields a maximum hole-boring velocity of $v_{hb}/c = d\lambda/2\lambda \cong 0.024$. Cold aluminum ions reflecting off this moving layer would gain a velocity of $2v_{hb}$ and energy of 29 MeV, which is 5 times lower than the measured 165 MeV ion peak.

The charge-to-mass ratios of $Al^{11+}$ (0.407) and $C^{5+}$ (0.417) are close. Therefore conceivably the $Al^{11+}$ trace in the TP data could be contaminated by $C^{5+}$ from hydrocarbon contamination of aluminum foils. To address that concern, we deliberately coated the rear side of 110 nm aluminum targets with a 10 nm of carbon layer to mimic such contamination and repeated the same experiment (Table 1 – fourth row). Fig. 3a shows the raw TP data from such experiment. The dominant trace is $C^{6+}$, not $C^{5+}$. This differs significantly from the pure aluminum foil result in Fig. 2a. The $C^{6+}$ trace (Fig. 3b) shows a spectral peak around 80 MeV (i.e., 6.7 MeV/nucleon) similar to the 6.2 MeV/nucleon $Al^{11+}$ ion spectral peak from a pure aluminum target. This suggests that the novel dynamics reported here, operating at the rear of the target, are robust and transferable to other ion species with proper optimization. The protons show no spectral peak above 9 MeV (Fig. 3c).

## Generation of 310 MeV $Al^{11+}$ & 120 MeV $C^{6+}$ ion beams

The Trident laser was focused with a faster f/1.5 off-axis-parabola, creating a peak laser intensity of $8 \times 10^{20}$ W/cm$^2$ (4× the f/3 intensity) onto a 250 nm thick aluminum foil (Table 1 – fifth row). Here, two newly developed high-dispersion TPs provide simultaneous on-axis and 11° off-axis ion spectral measurements (see methods and supplementary Fig. 4). The raw TP data (Fig. 3d) shows well-separated $Al^{11+}$, $Al^{12+}$, $Al^{13+}$ and proton traces. The $Al^{11+}$ trace is still dominant. The on-axis $Al^{11+}$ ion spectrum (red solid line in Fig. 3e) peaks at 310 MeV (11.5 MeV/nucleon) with 41% energy spread (total of 4.2 x $10^8$ ions/msr with average energy of 179 MeV). The simultaneous off-axis $Al^{11+}$ spectrum (dotted blue line in Fig. 3e) peaks at 250 MeV (9.3 MeV/nucleon) with 21% energy spread

(total of 4.7 x $10^8$ ions/msr with average energy of 167 MeV). The proton spectra (Fig. 3f) peaks around 12 MeV, a similar energy/nucleon as the $Al^{11+}$.

The ion/proton beam profile (Fig. 3g) shows the beam extending up to a 28° FWHM. Assuming radial symmetry, we estimate a total of 1 x $10^{11}$ $Al^{11+}$ ions in the full beam (average energy of 173 MeV). This implies a ~3J aluminum beam out of the 80J incident laser energy (4% conversion efficiency). A similar calculation for protons yields ~0.2% conversion efficiency. This interaction also exhibits plasma expansion towards the laser of 1.1 µm/ps derived from the reflected FROG measurement (Fig. 3h), and 1.2 µm/ps from the reflected spectral peak blue-shifted by 8 nm (Fig. 3i).

Once again, we address possible hydrocarbon contamination with targets by coating them with 10 nm of carbon layer on the rear side (Table 1 – sixth row). The raw TP data from these targets (Fig. 3j) again shows predominantly $C^{6+}$. The $C^{6+}$ spectrum (Fig. 3k) peaks around 120 MeV (10 MeV/nucleon) measured on-axis, and around 100 MeV (8.3 MeV/nucleon) measured off-axis, similar to $Al^{11+}$ energy/nucleon spectral peaks from pure aluminum targets. The proton spectrum from the carbon coated aluminum foil (Fig. 3l) peaks at 18 MeV (on-axis) and 12 MeV (off-axis).

## Generation of 220 MeV $C^{6+}$ ion beam

Next, we extend the narrow ion energy spreads from aluminum foils to synthetic diamond foils by reducing the diamond foil pre-expansion and mimicking its early expansion dynamics to match that of the aluminum foil discussed above. Contrary to the f/1.5–aluminum experiment (Table 1 – fifth row), a similar laser interaction with a 250 nm thick diamond foil (Table 1 – seventh row) generates exponential ion spectra for both $C^{6+}$ and protons (blue dotted line in Fig. 4b & 4c respectively). The early plasma dynamics of this interaction differs significantly from the aluminum case discussed above – the reflected light lacks both the initial frequency blue-shift (blue dotted line in Fig. 4e) and the blue spectral peak (dashed red line in Fig. 4d). Also, the reflected light peak is red-shifted by 4.5 nm, compared to only 1.7 nm for aluminum (solid red line in Fig. 3i). These results indicate again that the laser pedestal pre-expanded the diamond foil to a significantly lower peak plasma density than aluminum foil so that the main laser pulse can push the plasma forward almost right away.

In order to reduce the diamond foil pre-expansion, we simply reduced the incident laser energy from 80J to 60J which correspondingly reduced the pedestal and main pulse intensities by 25% (Table 1 – eighth row). The corresponding raw TP data is shown in Fig. 4a. The measured on-axis $C^{6+}$ ion spectrum (solid red line in Fig. 4b) is peaked at 220 MeV (i.e., 18.3 MeV/nucleon). The simultaneous off-axis spectrum (dashed green line in Fig. 4b) is peaked at 106 MeV (8.8 MeV/nucleon) and at 150 MeV (12.5 MeV/nucleon). The corresponding proton spectrum peaks at 23.3 MeV measured on-axis (solid red line in Fig. 4c) and 17.8 MeV measured off-axis (dashed green line in Fig. 4c). The same estimation as before yields a ~4% conversion efficiency (total of 2 x $10^{11}$ ions, 81 MeV average energy) for $C^{6+}$ and 0.6% for protons.

The reflected light from this reduced-intensity interaction shows a diamond foil expansion very similar to the aluminum foil interaction discussed earlier – early frequency blue-shift of $d\omega/\omega = 0.008$ (dotted blue line in Fig. 4f) i.e., a plasma expansion of 1.2 µm/ps towards the laser; 8 nm blue-spectral peak in the reflected light (solid blue line in Fig. 4d). Overall, these results indicate that maintaining higher initial plasma density by reducing the foil pre-expansion is a key to accessing the self-organization dynamics discussed earlier.

**Particle-In-Cell simulations**

2D Particle-In-Cell simulations using the Vector-Particle-In-Cell (VPIC) code [50] helped interpret the results. In the simulation, the laser parameters are: intensity=8 x $10^{20}$ W/cm$^2$; FWHM duration=650fs; full-duration=1400fs (i.e., 420 µm long in free-space). The plasma parameters are: 250 nm thick aluminum foil pre-expanded by the Trident laser pedestal per HELIOS code modelling (Fig. 5g inset); peak plasma density=250$n_{cr}$ located at x=95.4 µm; ion charge-to-mass ratio = 0.5 (see methods for further details). The laser time-markers are: laser launched from left boundary at 0fs; laser reaches target at 315fs; onset of transparency at 950fs; laser exits plasma at 1785fs; laser exits right boundary at 2100fs; simulations ends at 2258fs.

In the simulation, the initial laser-plasma dynamics and the onset of relativistic transparency proceed similarly to earlier reports in this regime [41]. After transparency, the laser drives a large forward electron current. Here, we focus primarily on the plasma dynamics once the laser exits the plasma, resulting in ion energy spread reduction. Fig. 5a shows self-generated azimuthal magnetic fields (up to 3 x $10^4$ Tesla) at $t$=1277fs (i.e., 327fs after transparency) by the forward and transverse electron currents. Within the plasma channel ($t$=1855fs), the laser keeps driving electrons forward against the magnetic resistance and sustaining a net negative charge density (Fig. 5b). The ions within the plasma channel exhibit an exponential spectrum (not shown) at this time.

As the laser exits the simulation box ($t$=2048fs), a strong azimuthal magnetic field surrounding the plasma channel persists (Fig. 5c and supplementary video 1 https://youtu.be/x1vdl1X8qoQ). Fig. 5d shows the persistent longitudinal electric field $E_x$ in the middle of the simulation box (marked by the white rectangle box in Fig. 5b) at the same time. The negative $E_x$ from $x$=30 μm to $x$=95μm injects the electrons from this region into the plasma channel partially replenishing the forward loss of electrons. The positive $E_x$ peak (Fig. 5d), coincident with a B-field kink (Fig. 5c), is the result of electron bunching at $x$ = 139μm in the absence of the driving laser (marked by a black circle in Fig. 5e).

The ions continue to move forward unaffected by the magnetic field leading to a net positive charge around the electron bunch (marked by black circle in Fig. 5e). Such a charge distribution rearranges itself by accelerating the (lagging) slower ions and decelerating the (leading) faster ions, creating an ion spectral and density peak at $x$ = 149 μm (marked by black circle in Fig. 5f) moving at 0.15$c$ (Fig. 5e and 5f). Fig. 5g (blue solid line) shows the final ion spectrum, peaked at 10.6 MeV/nucleon, consistent with the experiment. The electron and ion charge density snapshots in Figs. 5h and 5i, respectively, show the plasma channel and the electron/ion bunch at the end of the simulation. This simulation consumed 512,000 core-hours, at the extreme of our computational capability.

To check the effect of lower initial plasma density due to pedestal pre-expansion, we repeated the same simulation discussed above with a reduced peak density of $125n_{cr}$. In this case, plasma transparency occurred at 680 fs (270 fs earlier than the $250n_{cr}$ simulation). Once the laser exits the simulation box at $t$=2048 fs, the residual self-generated B-fields (Fig. 5j) and the longitudinal $E_x$ (not shown) are ~1/3 of those in the $250n_{cr}$ simulation, and the plasma channel is much wider (Fig. 5k). The ion spectrum at the end of the simulation does not show any pronounced peak (dashed black line in Fig. 5g), consistent with our experimental results.

## Discussion

The question arises as to whether the ion spectral peaking reported here could be due to a different mechanism, such as multispecies light-sail RPA in Ref.[14] or leaky light-sail regime RPA in Ref. [51]. In Ref. [14], the authors claim that their target remained highly reflective during the whole duration of the laser-plasma interaction. But in our case, we show both experimentally and computationally that the plasma becomes relativistically transparent near the peak of the main laser pulse thus significantly different from Ref. [14]. In leaky light-sail RPA [51], the lightest species (protons) exhibits a spectral peak, not the bulk, heavier species (C). In our work, Al & C foils (with proton contamination) show spectral peak in the bulk (and heaviest) species (see rows 2, 5, and 8 in table 1). Hence it is not conceivable that either of these mechanisms play a major role in our case. 'Magnetic vortex acceleration' is another mechanism that relies on self-generated quasistatic magnetic field at the rear side of the plasma for efficient forward ion acceleration and collimation [36-39]. In the magnetic vortex mechanism, magnetic pressure expels electrons and builds up an electrostatic field, which accelerates the ions forward at the plasma-vacuum interface[39]. In our case, we also see self-generated quasistatic magnetic field at the rear side of the plasma similar in nature to Ref. [39]. However, in our case the late-time dynamics responsible for the reduced energy spread of the ions (in the channel) depends on the slowdown and depletion of electrons injected from the front side of the plasma.

A second concern is whether protons racing ahead of heavier ions could tamp and spectrally bunch the heavier ions. We have studied such issue in the past both experimentally and computationally in a controlled fashion using thin carbon nanofoils either with proton contamination or said contamination eliminated by preheating the target [52]. The result was that the contaminant protons did indeed affect the carbon ion energy distribution, but it did so at the cutoff, i.e., energies in the 0.5-1 GeV (i.e., >42 MeV/nucleon equivalent to >1 GeV for Aluminum). Therefore, it does not seem reasonable that tamping of heavier ions by protons is dominant here, where the highest energy peaks are 310 MeV for Al and 220 MeV for C.

Another question is whether somehow different charge states of Al (i.e., $Al^{12+}$ and $Al^{13+}$) could race ahead of $Al^{11+}$ and tamp the $Al^{11+}$ energy distribution. In the case of $C^{6+}$, there is no higher C charge state to tamp the distribution; therefore at least in this case the concern is not pertinent. In the case of the f/3 laser focusing on Al (row 2 in Table 1), we have simulated the problem with $Al^{11+}$ (81.7%), $Al^{12+}$ (18.1%) and $Al^{13+}$ (0.2%) (Not shown here due to space constraints). The simulation showed no significant layering of the different Al charge states because their charge-to-mass ratios are very close to each other. It is therefore inconceivable that the $Al^{11+}$ energy distribution is dominated by tamping from $Al^{12+}$ and $Al^{13+}$.

Finally, we have observed proton spectral peaks with a similar energy/nucleon as the heavier bulk species in the f/1.5 case (rows 5, 6, and 8 in table 1) where it lies within the observable range of our TP detector. Although it is conceivable that this could be due to "buffered acceleration" involving relativistic transparency [53], there are some remaining concerns to make such a claim. In Ref. [53] only the lightest species (protons) showed a spectral peak while the heaviest species (C) energy spectrum was exponential. Moreover, the proton energy peak was at a much higher value than the average energy per nucleon of the bulk heavier species (C). We acknowledge that we do not have a clear explanation yet as to why we observe proton spectral peaks with a similar energy/nucleon as the heavier bulk species.

In summary, computer simulations show a self-organizing scheme that reduces the ion energy spread using self-generated fields from optimized laser-plasma interactions in the relativistic transparency regime. Consistent with simulations, we have demonstrated laser-driven ion beams with narrow spectral peaks at energies up to 18 MeV/nucleon and ~5% conversion efficiency from 0.12 PW laser interactions with planar foils. The energy spread and efficiency could be further improved with better target design, pre-expansion control and external magnetic fields. We also demonstrate that increasing the focused laser intensity fourfold (by reducing the focusing optic f/number twofold), increases the spectral-peak energy twofold. This implies that 33 MeV/nucleon carbon ion beam with narrow energy spread, as required for ion-fast ignition, could be generated using PW scale lasers. Furthermore divergence control (i.e., collimation/focusing) of these energetic narrow energy spread ion beams is a crucial next step in making these ion beams suitable for various applications[54-57].

Methods

**Laser system and ion diagnostics.**

The experiments were conducted at Trident laser facility at Los Alamos National Laboratory, USA. The Trident laser (80J, 650 fs FWHM, 1053 nm wavelength, linear polarization) is focused at normal incidence onto the target using an f/3 off-axis parabola (OAP) to a spot size of 10 µm diameter (first Airy minimum containing 65% laser energy) with a peak laser intensity of $2 \times 10^{20}$ W/cm$^2$ ($a_0 \approx 13$). Plasma mirrors were not used in the experiment. A high resolution TP employing 0.91 Tesla magnetic field over 20 cm long and a pair of copper electrodes, also 20 cm long, charged up to 28 kV potential was used to quantify the resulting ion spectra from the laser-plasma interaction at on-axis (0 degrees) and off-axis (8.5 degrees) [58]. Image plates were used as ion detectors in the TP and they were cross-calibrated against CR-39 nuclear track detectors. The instrumental cut-off is at 50 MeV for Aluminum ions due to an 18 µm thick aluminum foil placed in front of the ion detector to block laser light. The protons had a low energy cut-off at 9 MeV due to the size of the electrodes and the image plate. For this particular experiment the TP was rotated between on-axis and off-axis for ion spectra characterization. The divergence of the ion beam profile was characterized using a magnetic spectrometer called iWASP (ion wide angle spectrometer) [59] and image plate detectors. The iWASP was used only on selected shots as it would block the use of any diagnostic located further downstream.

The second set of experiments used a faster f/1.5 OAP to focus the same Trident laser to a spot size of 5 µm diameter (first Airy minimum containing 65% laser energy) with a peak laser intensity of $8 \times 10^{20}$ W/cm$^2$ ($a_0 \approx 26$). For f/1.5 experiments we developed two additional high-dispersion Thomson parabolas employing longer electrodes (up to 50 cm), shorter magnetic field (0.82 Tesla over 10 cm long), and longer drift length that enabled better separation of traces with different charge states. The two TPs were located side-by-side on-axis and 11° off-axis enabling the simultaneous on-axis and off-axis TP data collection (see supplementary figure 4 for more details). The maximum raw signal strength in the TP data reported in this article was $5.7 \times 10^4$, which is well below the saturation limit of $6.5 \times 10^4$. The ion beam profile in this experiment was measured using a radio chromic film (25 cm x 20 cm with 1 cm gap in the middle for downstream diagnostics) instead of the iWASP magnetic spectrometer used earlier. The front of the RCF was covered with 75 µm thick aluminum foil to block the laser light, aluminum ions up to 180 MeV and protons up to 3 MeV. Although we cannot separate the ion and proton beam profiles using the RCF signal in our configuration, we assume they have similar divergence as shown by the earlier iWASP measurements in Fig. 2e.

**Particle-In-Cell (PIC) simulation.**

Two-dimensional VPIC (Vector-Particle-In-Cell)[50] simulations were performed with the Trident laser parameters. The simulation box size is 200 µm in *x* direction (laser propagates from left to right) and 50 μm in *y* direction with an exponential aluminum plasma profile facing the incoming laser. In the simulation, the laser is focused at *x*=95.4 μm. The transverse laser profile is Gaussian with intensity spot size of 10 μm (5 μm) diameter and a peak intensity of 2 x $10^{20}$ W/cm$^2$ (8 x $10^{20}$ W/cm$^2$). The plasma profile is exponential with a peak density of $250n_{cr}$ or $125n_{cr}$, located at *x*=95.4μm. We used 382 (270) and 256 (180) cells per wavelength along *x* and *y* directions, respectively for $250n_{cr}$ ($125n_{cr}$) plasma. For each cell, we used 625 macro particles for electrons and ions each (total of 8 x $10^9$ macro particles for electrons and ions each). Ion charge-to-mass ratio was kept at 0.5. Initial electron and ion temperatures were 32keV and 1keV, respectively. The laser is *p*-polarized in the simulation plane.

**Acknowledgements**

We gratefully acknowledge the support of the Los Alamos National Laboratory LDRD (Laboratory Directed Research and Development) program for this work. We also thank the capable and dedicated Trident laser personnel: R. P. Johnson, T. Shimada, R, Gonzales, S. Reid and R. Mortensen for the laser operation. The simulations were run using LANL Institutional Computing and ASC Capability Computing Campaign allocations. We thank L.Yin and B. Albright for their help with VPIC simulations. S.P thanks J. Schreiber for the thoughtful discussions.

**Author contributions**

S.P, D.C.G, R.C.S and J.C.F designed and executed the experiment. C.E.H, M.A.S and C.K made the targets. S.P analyzed the experimental data. C.H performed the VPIC simulations. S.P wrote the paper with contributions from all the co-authors.

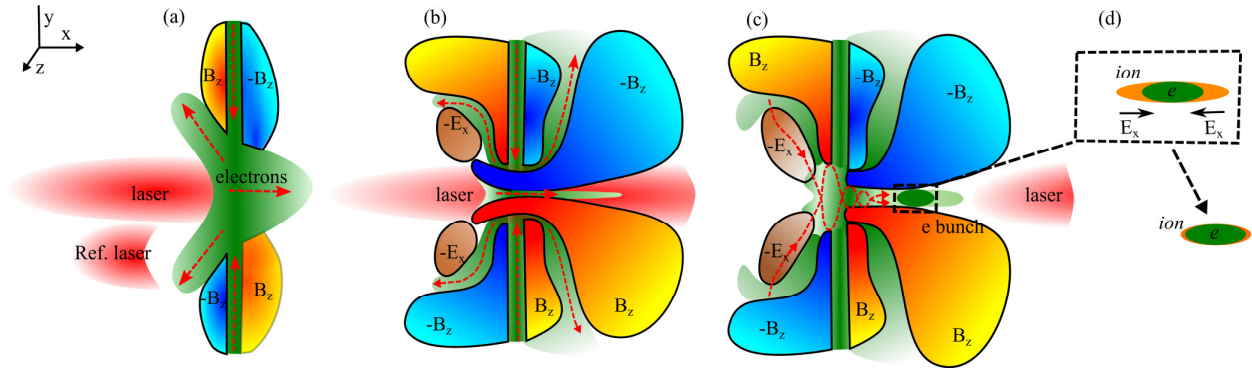

**Figure 1: Schematic of the three phases of laser-plasma dynamics leading to a narrow energy spread ion beam generation via self-generated plasma fields.** (**a**) TNSA-like phase in opaque plasma; (**b**) relativistic transparency and ion acceleration phase; (**c**) energy-spread reduction phase occurring after the laser exits the plasma; (**d**) enlarged view of the boxed region in panel (c) showing the ion spectral and spatial bunching.

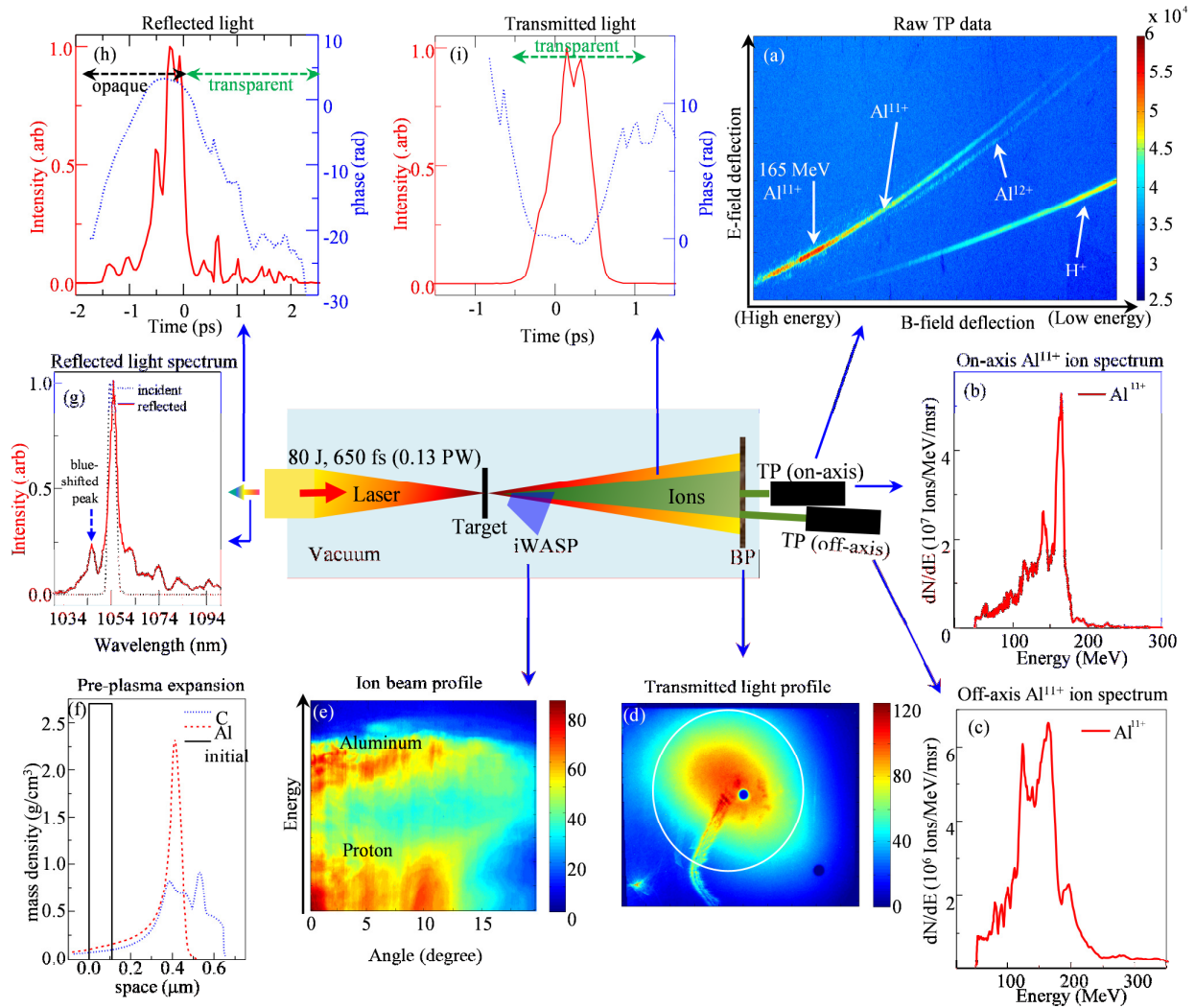

**Figure 2: Schematic of experimental setup (f/3 laser focus onto 110 nm Al foil).** The 0.12 PW Trident laser is focused with f/3 off-axis-parabola (peak intensity ~ 2 x $10^{20}$ W/cm$^2$) onto a 110 nm thick aluminum foil (Table 1 – second row). **(a)** raw Thomson parabola (TP) data; **(b)** measured Al$^{11+}$ ion spectrum on-axis (solid red line); **(c)** measured Al$^{11+}$ ion spectrum 8.5° off-axis (solid red line). TP used image plate (IP) detectors cross-calibrated against CR-39 nuclear track detectors. The TP instrumental cut-off is at 50 MeV due to an 18 μm thick aluminum foil placed in front of the ion detector to block laser light; **(d)** transmitted laser beam profile captured on a Macor® plate (30.5cm x 30.5 cm square) and imaged onto a separate CCD camera (Apogee Alta U8300), which is used to quantify the amount of laser energy transmitted through the plasma; **(e)** angularly resolved ion energy spectra measured using an ion wide-angle magnetic spectrometer (iWASP); **(f)** simulation result of

Trident laser pedestal pre-expansion of 110 nm aluminum and diamond foils using the HELIOS rad-hydro code; **(g)** reflected light spectrum (solid red line) measured using an infra-red spectrometer (Bruker Optics with Andor iDUS 1.7μm InGaAs CCD) along with incident laser spectrum (dotted black line); **(h)** time-resolved reflected light intensity (solid red line) and its temporal phase (dotted blue line) measured using a single-shot Frequency-Resolved-Optical-Gating (FROG). The positive (negative) slope of the temporal phases indicate spectral blue (red) shift via $\omega_t = d\phi/dt$, where $\omega_t$ is the instantaneous angular frequency. **(i)** time-resolved transmitted light intensity (solid red line) and its temporal phase (dotted blue line) measured using a separate FROG.

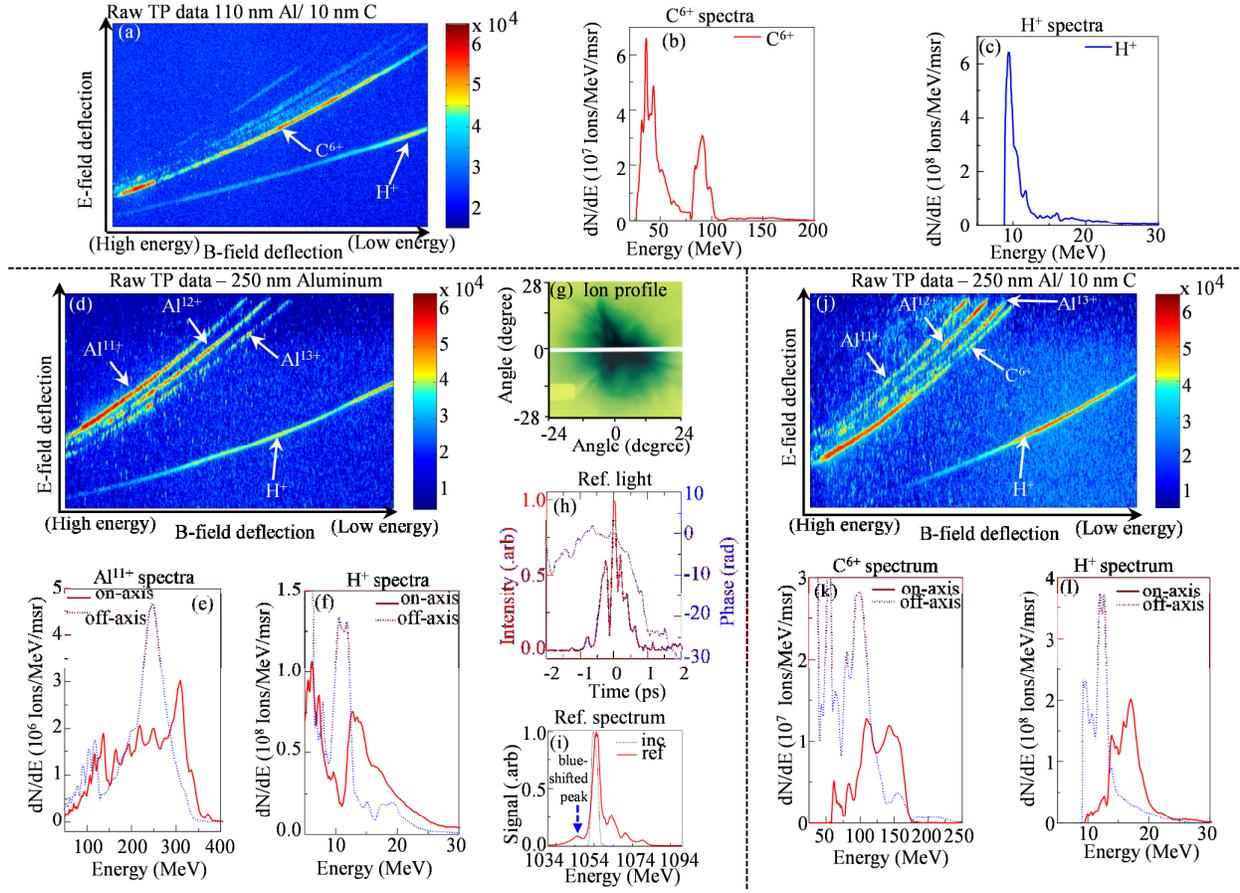

**Figure 3: Results from f/3 laser focus interaction with 10 nm carbon coated 110 nm Al foil (a-c) (Table 1 – fourth row) & f/1.5 laser focus onto 250 nm Al foil (d-i) (Table 1 – fifth row), and the same interaction with 10 nm carbon coated on the rear-side of 250 nm Al foil (j-l) (Table 1 – sixth row).** (a) Raw TP data from 110nm Al/10nm C foil showing pre-dominant $C^{6+}$ trace; (b) measured on-axis $C^{6+}$ ion spectrum (solid red line); and (c) measured on-axis proton spectrum (solid blue line). (d) raw TP data from the 250 nm Al foil interaction; (e) $Al^{11+}$ spectra on-axis (solid red line) and 11° off-axis (dotted blue line); (f) proton spectra on-axis (solid red line) and 11° off-axis (dotted blue line); (g) ion/proton beam profile measured using radiochromic film (RCF); (h) time-resolved reflected light intensity (solid red line) and temporal phase (dotted blue line); (i) time-integrated reflected light spectrum (solid red line) along with incident light spectrum (dotted black line). (j) raw TP data from the 10 nm carbon coated 250 nm thick Al foil; (k) $C^{6+}$ spectra measured on-axis (solid red line) and 11° off-axis (dotted blue line); (l) corresponding proton spectra measured on-axis (solid red line) and 11° off-axis (dotted blue line)

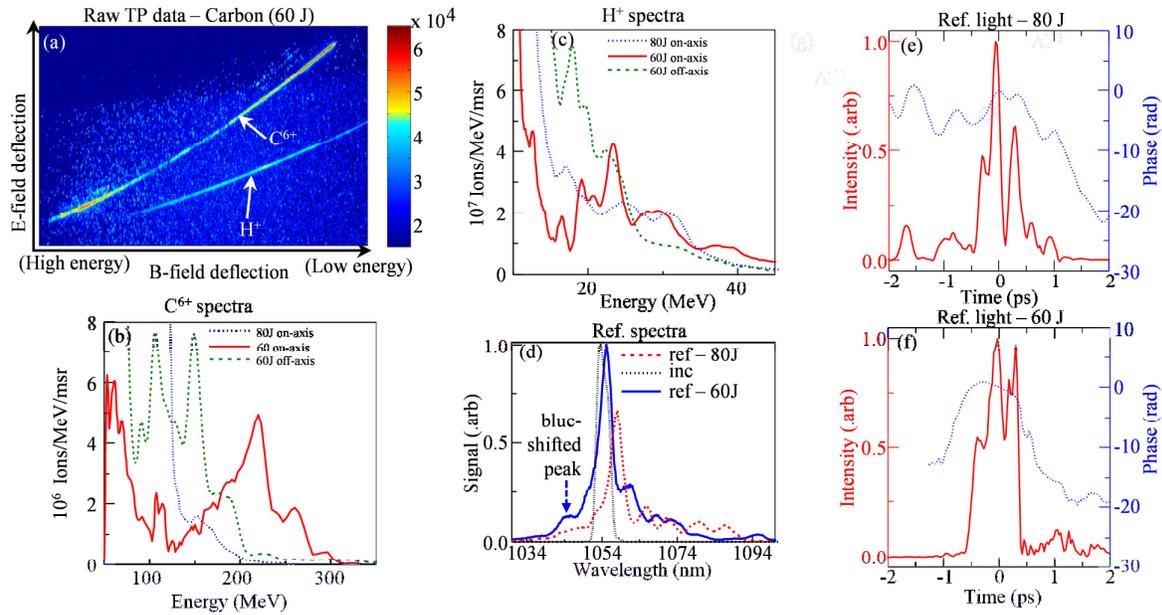

**Figure 4: Results from f/1.5 laser focus onto 250 nm synthetic diamond foil (Table 1 – seventh and eighth rows).** **(a)** raw TP data from 60J laser shot onto 250 nm diamond foil; **(b)** on-axis $C^{6+}$ spectrum (solid red line) and off-axis $C^{6+}$ spectrum (dashed green line) from 60J shot. On-axis $C^{6+}$ spectrum (dotted blue line) from 80J shot; **(c)** corresponding on-axis $H^+$ spectrum (solid red line) and off-axis $H^+$ spectrum (dashed green line) from 60J shot. On-axis $H^+$ spectrum (dotted blue line) from 80J shot; **(d)** Reflected light spectrum from 80J shot (dashed red line) and from 60J shot (solid blue line) along with incident laser spectrum (dotted black line); **(e)** time-resolved reflected light intensity (solid red line) and temporal phase (dotted blue line) from 80J shot; **(f)** time-resolved reflected light intensity (solid red line) and temporal phase (dotted blue line) from 60J shot.

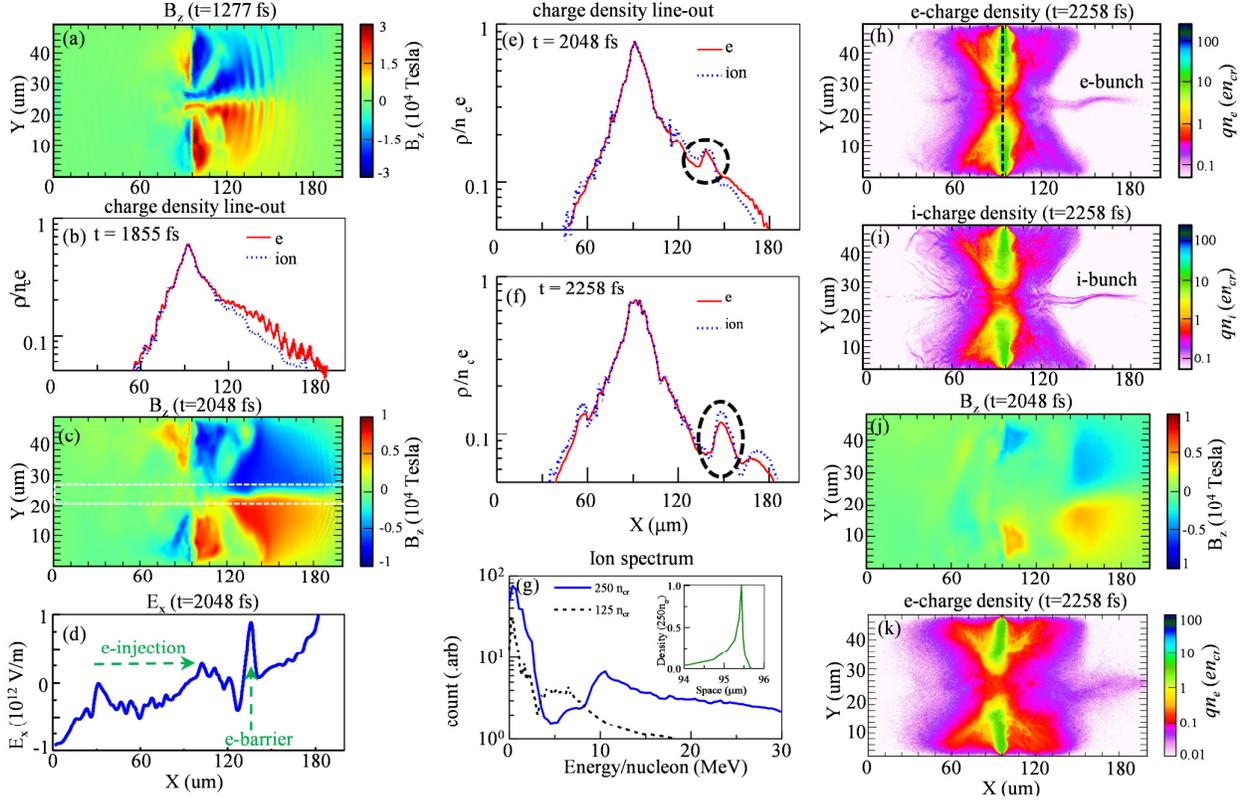

**Figure 5: 2D VPIC simulation results from f/1.5 laser focus onto pre-expanded 250 nm thick aluminum foil with peak plasma density of $250n_{cr}$ and $125\ n_{cr}$.** (a) Self-generated azimuthal magnetic field ($B_z$) from $250n_{cr}$ plasma at $t$=1277fs. Plasma transparency occurs at 950fs; (b) the electron and ion charge density line-outs along the plasma channel at $t$=1855fs 2048fs averaged over $y$=21μm to $y$=27μm; (c) self-generated azimuthal magnetic field ($B_z$) at a later time $t$=2048 fs when the laser exits the right boundary of the simulation box; (d) line-out of the longitudinal electric field at the same time $t$=2048fs averaged over $y$=21μm to $y$=27μm (white rectangle box in Fig. c). Dashed green arrows shows the electron injection and localization regions; **(e&f)** show the electron and ion charge density line-outs along the plasma channel at $t$=2048fs, and $t$=2258 fs respectively. The localized electron/ion population (marked in dashed black circle) moves at 0.15$c$ speed; (g) final ion spectrum from 250 $n_{cr}$ plasma (solid blue line) and from 125 $n_{cr}$ plasma (black dashed line) averaged over $y$=21μm to $y$=27μm, inset shows initial plasma profile for 250 $n_{cr}$ plasma used in the simulation; (h) final plasma electron charge density profile from 250 $n_{cr}$ plasma at the end of the simulation at t = 2258fs. Dashed black line shows the size and location of the initial plasma; (i) corresponding plasma ion charge density profile; (j) self-generated azimuthal magnetic field ($B_z$) from 125 $n_{cr}$ plasma at t= 2048 fs when the laser exits the right boundary of the simulation box; (k) final plasma electron charge density profile from 125 $n_{cr}$ plasma at the end of the simulation at t = 2258fs.

# Supplementary information for "Efficient quasi-monoenergetic ion beams up to 18 MeV/nucleon via self-generated plasma fields in relativistic laser plasmas"

Sasi Palaniyappan[*,1], Chengkun Huang[1], Donald C. Gautier[1], Christopher E. Hamilton[1], Miguel A. Santiago[1], Christian Kreuzer[2], Rahul C. Shah[1], and Juan C. Fernández[1].

[1]*Los Alamos National Laboratory, Los Alamos, New Mexico 87545, USA.*
[2]*Ludwig-Maximilian-University, Munich, Germany.*

*e-mail:* sasi@lanl.gov

**Evolution of self-generated plasma magnetic field**

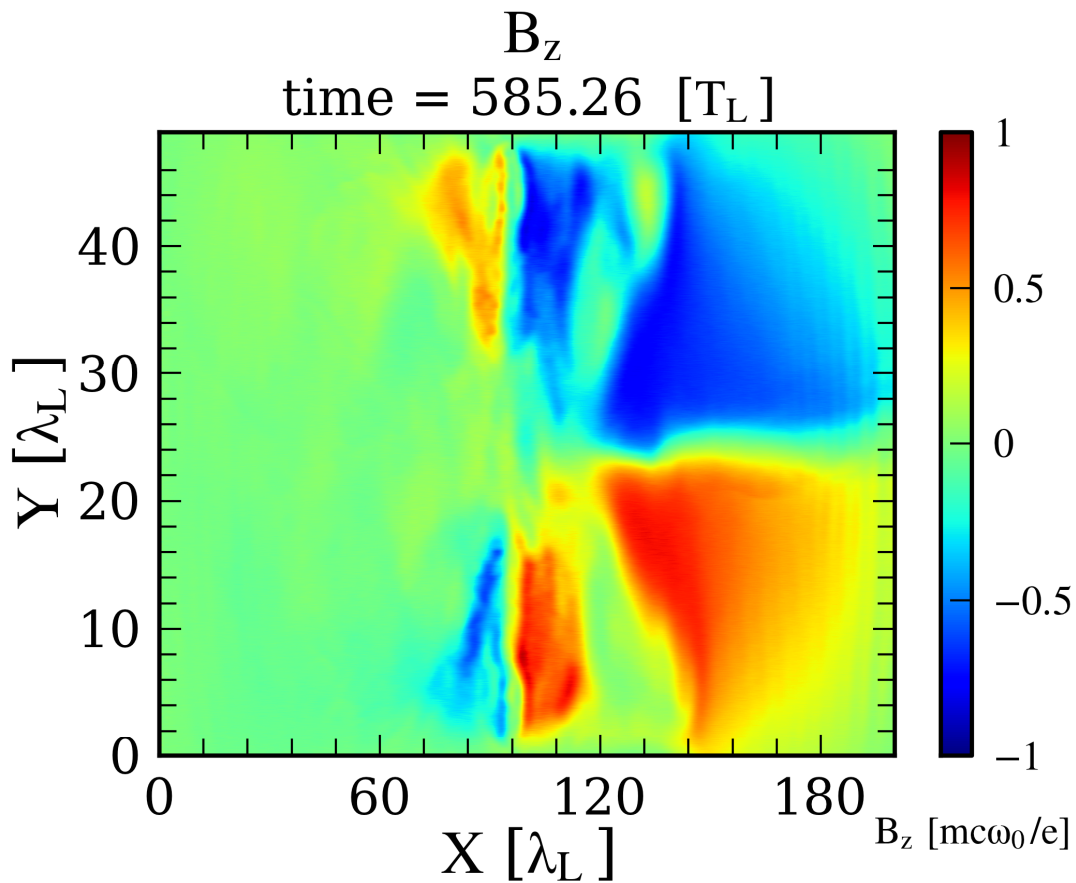

**Supplementary Video 1: Evolution of self-generated plasma magnetic field.**

Supplementary video 1 shows the evolution of self-generated plasma magnetic field ($B_z$) from the VPIC simulation with Trident laser interacting with $250 n_{cr}$ plasma. The laser has a Full-

Width-Half-Maximum of 650 fs and full duration of 1400 fs (i.e., 400 optical cycles). The time-stamp in the video is in units of laser optical cycle i.e., 3.5 fs. The laser is launched 5 microns away from the left boundary at $t=0$ fs. Coming from the left boundary, the laser takes 315 fs (~90 optical cycles) to reach the target placed at 95 μm. At t = 1785fs (~510 optical cycles) the trailing edge of the laser begins to exit the plasma. The simulation ends at t=2258fs (645 optical cycles)

**Atomic ionization of aluminum**

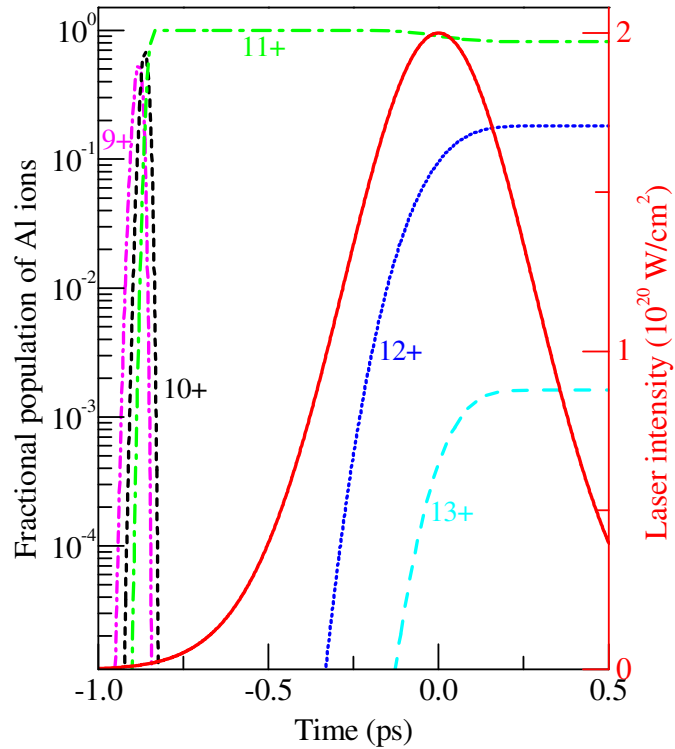

**Supplementary Figure1: Laser field ionization of aluminum atom by the Trident laser pulse (f/3 focus).**

Supplementary figure 1 shows the fractional population of aluminum ions when Trident laser pulse (f/3 focus) ionizes an aluminum atom. This is calculated using Ammosov-Delone-Krainov (ADK) tunneling ionization rate (Ammosov, M.V., N.B. Delone, and V.P. Krainov, Sov. Phys. JETP, 1986. **64**: p. 1191-1194). The result shows that almost the entire aluminum atom becomes aluminum 11+ ion at the foot of the laser pulse ( the ionization potential to strip an electron from $Al^{10+}$ to get $Al^{11+}$ is 442 eV). The populations of aluminum 12+ and 13+ ions begin to increase only near the peak of the laser pulse due to larger *K*-shell ionization gap of 1644 eV (the ionization potential to strip an electron from $Al^{11+}$ to get $Al^{12+}$ is 2086 eV). The populations of $Al^{11+}$, $Al^{12+}$ and $Al^{13+}$ ions remain at a fixed level of 81.7%, 18.1% and 0.2% for most of the laser pulse duration. Since the peak intensity is reached only at the middle of the focus, in the Thomson parabola we should expect to see mostly $Al^{11+}$ ions along with very little $Al^{12+}$ from the Trident laser pulse interaction with aluminum.

**Proton Spectra**

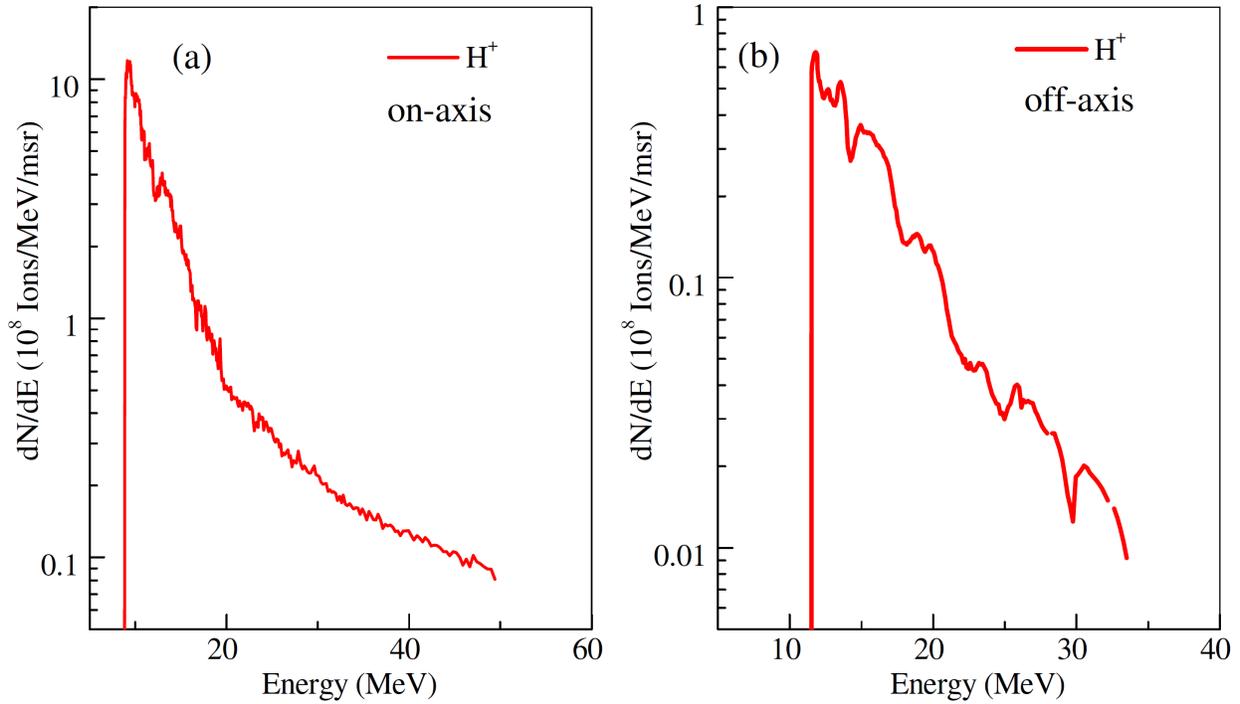

**Supplementary Figure 2: Proton energy spectra.**

Supplementary figure 2 (a&b) show the proton energy spectra from Trident laser (peak intensity $2 \times 10^{20}$ W/cm$^2$) interacting with 110 nm thick aluminum foil measured on-axis (a) and 8.5° off-axis (b). The spectral shape is exponential above the instrumental cut-off at 9 MeV. Calculation shows $4 \times 10^9$ protons per millisteradian on-axis and $2.8 \times 10^8$ protons per millisteradian off-axis with average energy of 13 MeV. Estimation of the total number of protons with a methodology similar to the one used for $Al^{11+}$ in the manuscript yields a total of $2.3 \times 10^{11}$ protons corresponding to proton conversion efficiency of 0.6%.

**Trident Laser Contrast**

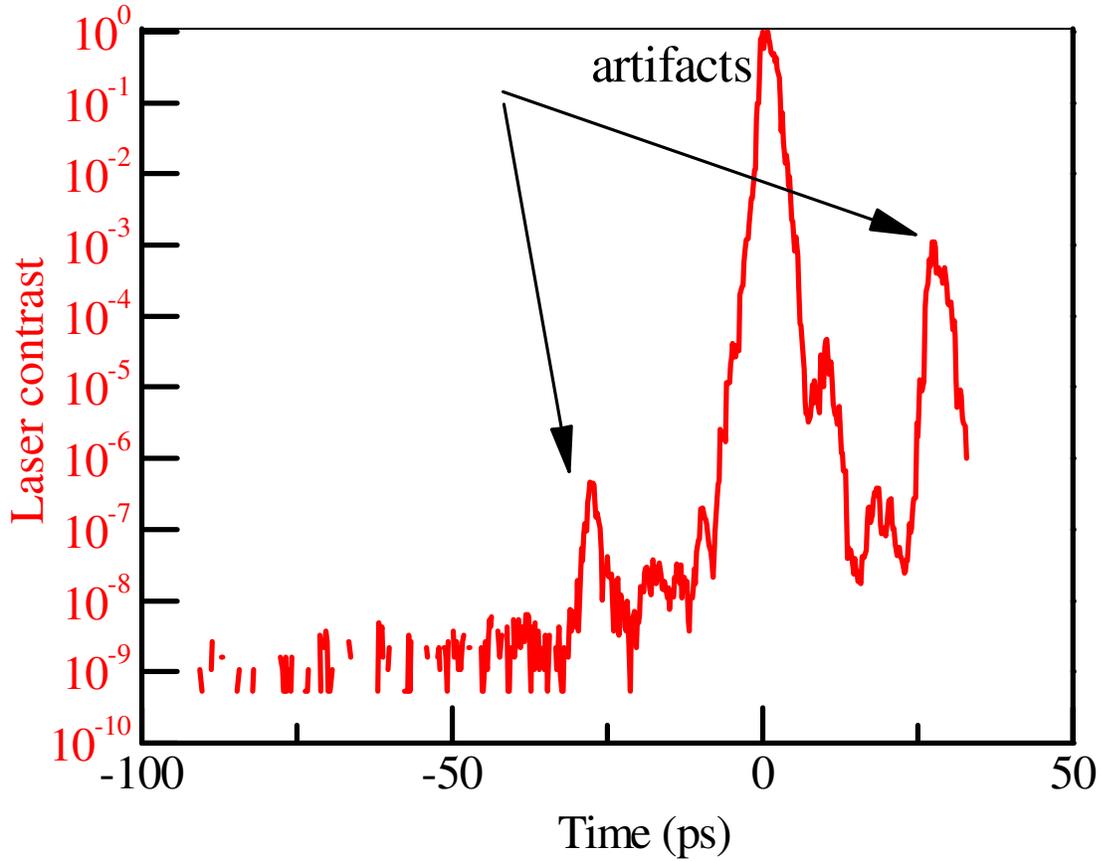

**Supplementary Figure 3: Trident laser contrast.**

Supplementary figure 3 shows the Trident laser contrast measured using a scanning third order auto-correlator (Rincon, DelMar Photonics). The laser pedestal drops to $10^{-8}$ nearly 10 ps before the laser peak. The detection limit is $10^{-9}$. The two peaks at ±27 ps are measurement artifacts from the scanning device.

**High-dispersion Thomson Parabola**

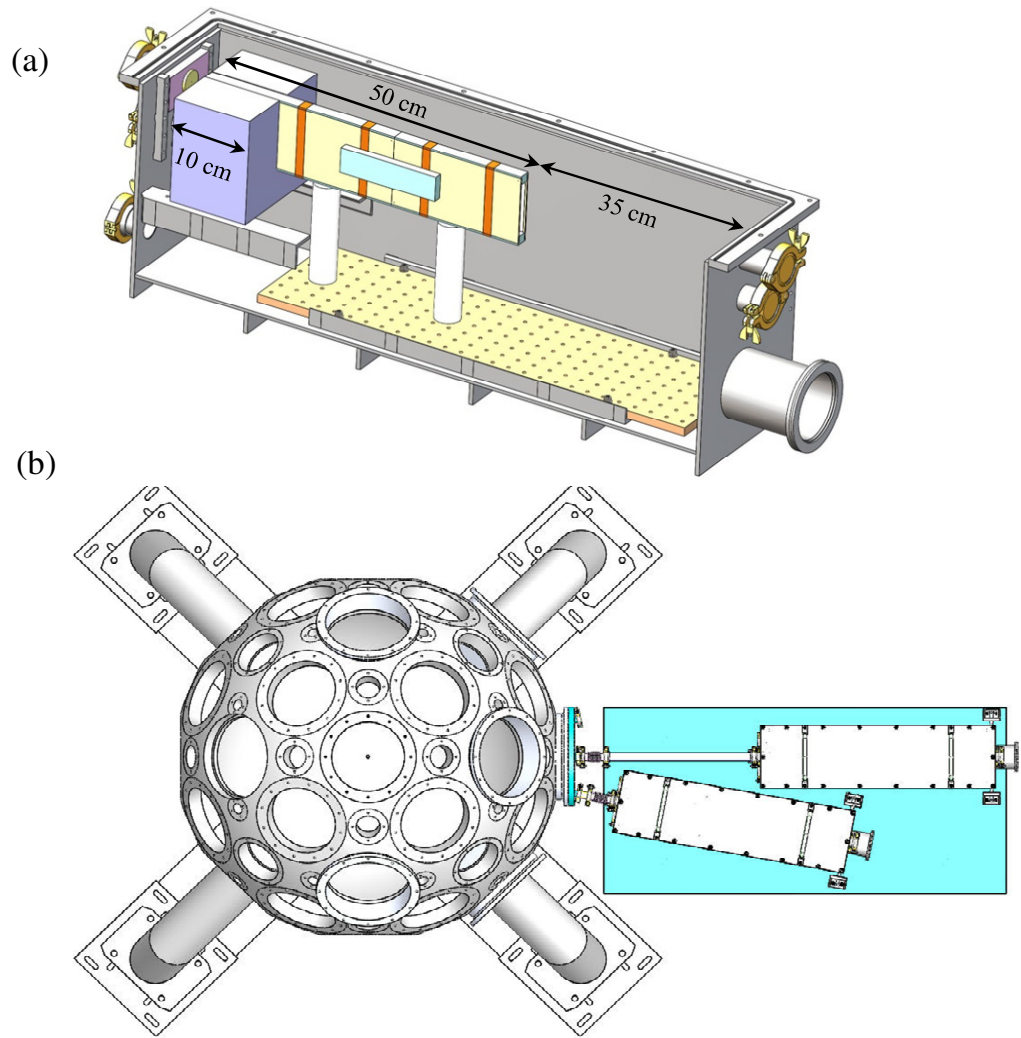

**Supplementary Figure 4: (a) side-view of the high-dispersion Thomson parabola (TP) ion spectrometer. (b) Two TPs stacked side-by-side to measure ion spectra on-axis and 11° off-axis.**

Supplementary figure 4a shows the side-view of the high-dispersion Thomson parabola without the front panel for better visibility. The TP uses 0.82 Tesla magnets (10 cm long) and 50 cm long electrodes that can be charged up to 15kV potential. The drift length from the end of the electrode to detector (placed at the end of the box) is 35 cm. Supplementary figure 4b shows the two TPs stacked side-by-side next to the Trident north target chamber. The two TPs simultaneously measured the on-axis and 11° off-axis ion spectra.